\documentclass[10pt,conference]{IEEEtran}
\IEEEoverridecommandlockouts
\usepackage[T1]{fontenc}
\usepackage[utf8]{inputenc}
\usepackage{cite}
\usepackage{amsmath,amssymb,amsfonts}
\usepackage{algorithmic}
\usepackage{graphicx}
\usepackage{textcomp}
\usepackage{xcolor}
\def\BibTeX{{\rm B\kern-.05em{\sc i\kern-.025em b}\kern-.08em
    T\kern-.1667em\lower.7ex\hbox{E}\kern-.125emX}}

\usepackage{pbalance}
\usepackage{subcaption}

\makeatletter
\newcommand{\linebreakand}{%
  \end{@IEEEauthorhalign}
  \hfill\mbox{}\par
  \mbox{}\hfill\begin{@IEEEauthorhalign}
}
\makeatother
\begin{document}

\title{Implementation of Dynamic Microsegmentation for a Multi-Robot Network}

\author{
\IEEEauthorblockN{\normalsize Vishwam Raval$^{\dagger}$}
\IEEEauthorblockA{\small \textit{Global Cyber Research Institute}\\
\textit{Texas A\&M University}\\
College Station, USA\\
vishwam@tamu.edu}
\and
\IEEEauthorblockN{\normalsize Jaewon Kim$^{\dagger \ast}$}
\IEEEauthorblockA{\small \textit{Global Cyber Research Institute}\\
\textit{Texas A\&M University}\\
College Station, USA\\
j1k@tamu.edu}
\and
\IEEEauthorblockN{\normalsize Ravi Pattabhi}
\IEEEauthorblockA{\small \textit{ColorTokens Inc}\\
San Jose, CA\\
ravi.pattabhi@colortokens.com}
\and
\IEEEauthorblockN{\normalsize Sandip Roy}
\IEEEauthorblockA{\small \textit{Global Cyber Research Institute}\\
\textit{Texas A\&M University}\\
College Station, USA\\
sandip@tamu.edu}
\thanks{$^{\dagger}$These authors contributed equally to this work.}
\thanks{This paper has been accepted for publication at the 2026 IEEE International Conference on Cyber Security and Resilience (CSR). Copyright 2026 IEEE.}
\thanks{$^{\ast}$Corresponding author: Jaewon Kim (j1k@tamu.edu).}
}

\maketitle

\begin{abstract}
This report describes a laboratory instantiation of a dynamic microsegmentation function for an Internet-of-Things network, specifically a multi-robot network, using a commercial software-defined microsegmentation tool.  Fine-grained dynamic segmentation policies are defined, where segments are adapted based on asset functional and health tags.  Automated implementation of the dynamic policies in software is also undertaken.  The microsegmentation solution is shown to automatically isolate an indicated threat in real time, while maintaining unaffected operations.  The demonstration indicates that microsegmentation can be implemented in IoT networks, in a way that allows for robust operation in the face of threats and disruptions.  
\end{abstract}


% \newpage
\section{Introduction}
Internet of Things (IoT) devices in industrial, defense, and critical-infrastructure environments have introduced substantial operational capabilities, with the global IoT device population projected to exceed 29 billion connected endpoints by 2030 \cite{iot1}. However, this growing dependence on IoT devices has also increased the susceptibility of critical systems to cyber threats, both by expanding the attack surface and by widening the potential impact radius of threats \cite{iot2}. Embedded endpoint devices frequently operate with constrained computational resources and limited onboard security, making them attractive initial entry points for adversaries \cite{iot3}. Once a single device is compromised, the open network architectures common in these environments allow attackers to move laterally, disrupt command hierarchies, manipulate sensing or actuation pipelines, and ultimately compromise the physical-world responses of multiple assets \cite{iot4}.

Traditional perimeter defenses are often ineffective in these dynamic and distributed settings \cite{iot5}. Embedded endpoint devices do not have the resources or operational flexibility to enact strong authentication or firewall systems; once a device is compromised, there is little to prevent it from launching attacks against controllers or neighboring devices, since it is within the trusted zone. Such vulnerabilities underscore the need for defense-in-depth or Zero-Trust Architecture (ZTA) principles for security, in which no device is inherently trusted and each communication must be continuously validated.

Network microsegmentation --  wherein tailored and dynamic/adaptive network security zones are used to limit lateral movement of network attacks -- forms a core component of a ZTA. There has been substantial adoption of microsegmentation in enterprise networks, with an estimated total market of about \$21 billion in 2025.  Microsegmentation solutions have proved effective in limiting lateral movement, and hence limiting risk, in the enterprise space. 

While there is a recognition that ZTA and specifically microsegmentation are important technologies for IoT and other cyber-physical systems, their deployment for these systems remains limited.  Microsegmentation for IoT networks entails several specific technical challenges: 

\noindent 1) Instantiation of dynamic segmentation rules in real-time for IoT endpoints, given computational constraints, heterogeneity of devices/protocols, and potentially ad hoc connectivity.

\noindent 2)  Detection of anomalies and classification of compromised assets, based on heterogeneous device-level, network-, and physical-world- behavior.

\noindent 3) Design of dynamic segmentation rules or policies, that maintain operational safety and enhance system resilience. 

This work presents an empirical investigation of dynamic microsegmentation for networked IoT devices, specifically a multi-robot network.  The work represents a collaboration between ColorTokens, a leading industry microsegmentation vendor with a growing IoT/OT customer base; and the Global Cyber Research Institute (GCRI) at Texas A\&M University, which has a research focus on cyber-physical and IoT security.  Using a controlled testbed at the GCRI, we have demonstrated a dynamic network segmentation solution that monitors asset cyber-physical behavior, classifies devices by operational and security state, and autonomously isolates compromised robots. Our architecture leverages tag-based asset health classification and API-driven policy orchestration, for context-aware reassignment of devices to restrictive quarantine segments immediately upon detection of malicious behavior.

The primary contribution of this study is the initial practical instantiation of an automated detection-response mechanism that integrates real-time asset health tracking with dynamic network segmentation.  This report describes results of the experiment, which demonstrate that compromised nodes can be isolated with minimal delay and without disrupting the availability or communication link of healthy robots. We believe that these findings represent a practical starting point for securing myriad IoT and cyber-physical systems, including autonomous vehicles, distributed industrial sensor networks, and other emerging Industry 4.0 infrastructures.

%Unlike static segmentation policies that require manual intervention, our system enforces zero-trust boundaries that adapt to changing threat conditions and are aware of operational context. 

% \clearpage
\section{Related Work}
Security challenges for IoT devices and CPS have received growing attention, as interconnected environments expose new vectors for lateral movement attacks, which can have severe physical-world consequences. Prior studies show that compromised edge devices can become effective entry points, allowing adversarial penetration and cascading failures across tightly coupled industrial control systems \cite{b1}.

Although IoT/CPS security and ZTA are both vibrant research areas \cite{zta1,zta2,cps1,cps2}, the intersection between the two is relatively new and unexplored.  In particular, only a limited set of studies have considered architecting microsegmentation for IoT networks \cite{b2,inter1,inter2,inter3}. 
These studies are primarily focused on developing frameworks and methods for microsegmentation, rather than operational implementations.  The study \cite{inter1} further develops a machine-learning-based approach for dynamic segmentation which is aligned with our effort, however experiments are limited to analysis of historical data sets and simulations rather than field demonstration.  

%adaptive research on Industrial Internet of Things (IIoT) segmentation frameworks underscores the importance of fine-grained isolation mechanisms for reducing the attack surface and securing critical operational subsystems \cite{b2}. 
There is a broader literature on intrusion detection for IoT and CPS systems (e.g. \cite{intrud1,intrud2,intrud3,b3,b4,b5,b6,ourwork1,ourwork2}) which is germane to this study, as identifying malicious or anomalous behavior is a prerequisite for timely isolation and containment. This literature spans a number of directions, but has a heavy focus on developing machine-learning and AI approaches for detection, including particularly distributed learning (e.g. federated learning) approaches.  Our group has also developed intrusion detection techniques \cite{ourwork1,ourwork2}, which aim to use physical-world response signals to build more generic detectors and also support more resource-constrained and/or mobile networks.

Policy automation and metadata-based network orchestration have also been explored as enablers of dynamic security enforcement in IoT environments. However, these frameworks assume the availability of powerful orchestration and stable connectivity. Such conditions are rarely present in distributed robotic systems, where nodes may be mobile, intermittently connected, or limited in computing resources \cite{b7}. Distributed robotic and cyber-physical systems frequently operate under mobility, intermittent connectivity, and resource constraints, limiting their ability to rely on centralized or cloud-assisted orchestration services \cite{b8}. As a result, existing automation techniques may not be readily applicable to CPS deployments that require rapid, fine-grained, and low-overhead security reconfiguration in response to evolving operational or threat conditions.

Together, the recent literature reveals several unresolved gaps. First, segmentation research does not address real-time dynamic microsegmentation in IoT and mobile/robotic systems. Second, advances in detection methodologies are generally not paired with automated containment mechanisms, leaving a gap between identifying compromised devices and preventing their lateral movement. Third, existing policy automation frameworks may lack the architectural flexibility and low-latency responsiveness required for IoT/robotic deployments. 

%Therefore, an end-to-end detection-response framework that unifies detection, dynamic re-classification, automated quarantine, and safe re-integration remains largely absent from the state of the art.  !!I like this sentence, but I'm worried that reviewers may state that we're not doing this also, so I'd rather be a bit defensive.

% \clearpage
\section{Experiment Setup}
\begin{figure}[h]
    \centering
    \includegraphics[width=0.99\columnwidth]{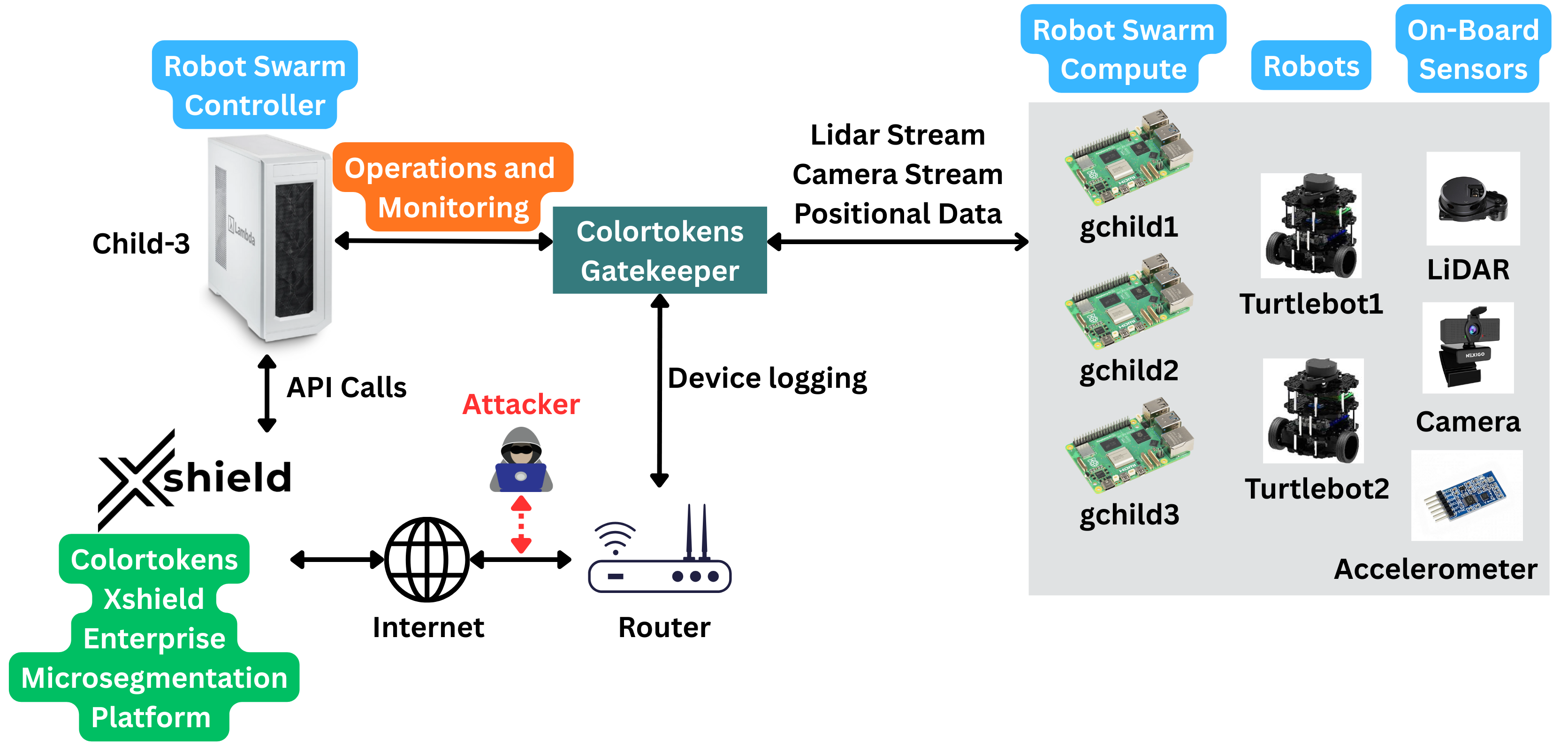}
    \caption{Configuration of Zero-Trust Microsegmentation and Connectivity for Networked Robot Application.}
    \label{fig:1}
\end{figure}
The experimental environment instantiates a multi-robot network characterized by a central command unit and distributed autonomous nodes. As shown in Fig.~\ref{fig:1}, the physical testbed consists of a Master Controller, three Turtlebots with embedded Raspberry Pi units functioning as autonomous robot nodes (designated as `\texttt{gchild1}', `\texttt{gchild2}', `\texttt{gchild3}'), and peripheral sensor arrays including LiDAR and cameras.

\begin{figure}[h]
    \centering
    \includegraphics[width=0.45\columnwidth]{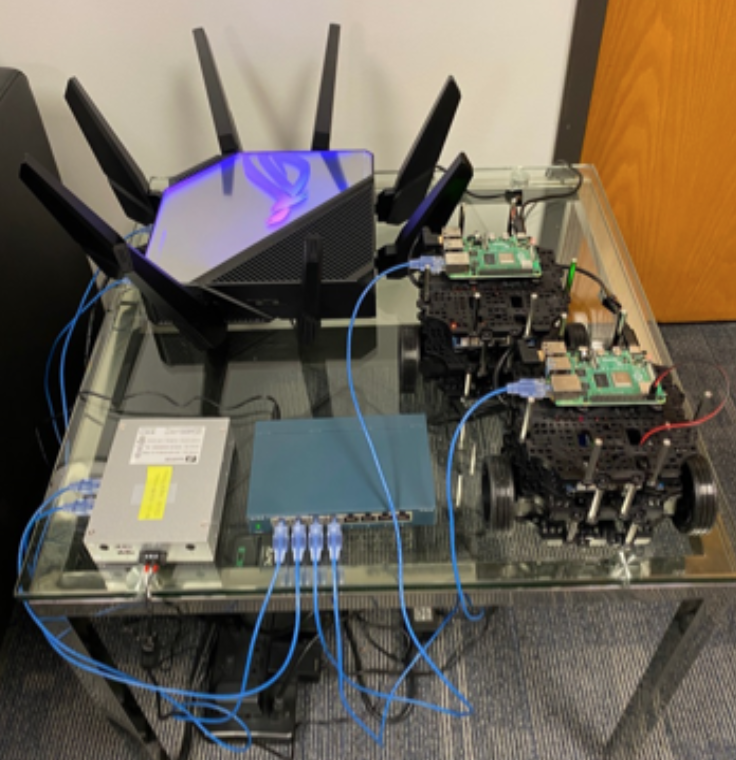}    
    \caption{Network Router, Network Switch, and ColorTokens Microsegmentation device connected to the Raspberry Pi-powered Robots.}
    \label{fig:1-1}
\end{figure}

The network setup for the experiment is as follows. A local area network for the robot team is formed using a gaming router (ASUS ROG Rapture WiFi 6E Router GT-AXE16000), which is also tied to the Texas A\&M network and hence the Internet via an Ethernet cable.  A network switch
(TP-Link 8 Port Gigabit Switch TL-SG108) is connected to the private side of the router. The Raspberry Pis on board the robots communicate via the switch.  A ColorTokens Gatekeeper, required for microsegmentation, is also connected to the private side of the network.  The gatekeeper has the ability to apply policies to all network traffic using the router.  The Gatekeeper is linked to the ColorTokens Xshield cloud platform via the router, where administrative credentials and API keys are configured to support device onboarding remote management, and segmentation policies are managed. In our setup, a ColorTokens dashboard for communication with the Xshield server is accessed through the Master Controller (a GCRI lab desktop), which also is connected to the router through the switch. The components are shown in Figure \ref{fig:1-1}.

We note that the ColorTokens microsegmentation function is designed to make network security as granular as possible by dividing the network into isolated segments, through the application of (possibly dynamic/adaptive) policies using the Gatekeeper. This allows for meticulous monitoring and control of traffic, with the goal of preventing unauthorized lateral movement within the network. With the ColorTokens microsegmentation function, security engineers can create secure zones to isolate environments, data centers, applications, and workloads across on-premises, cloud, and hybrid network environments, enhancing overall security posture.
While this experiment used a single Gatekeeper, we stress that network traffic across a wider footprint can be monitored/controlled in a coordinated way using multiple Gatekeepers, including in a High Availability network architecture model.

%To enforce network-level microsegmentation, a ColorTokens Gatekeeper was deployed at the network edge. This gateway functioned as the primary enforcement point, routing all traffic from the IoT and robotic assets to ensure comprehensive visibility and policy application. 

\section{Methodology}
This section outlines the experimental design, including the logical microsegmentation policies, and the automated enforcement algorithms used to validate the proposed zero-trust framework.

\subsection{Logical Segmentation and Policy Design}
To establish a robust Zero-Trust Architecture (ZTA), the network architecture abstracts away from rigid physical topology, relying instead on logical asset classification for access control. Each device within the experimental testbed is assigned specific metadata tags: an \textit{Application Tag} to denote its functional role (e.g., `Master' or `Robots') and a binary \textit{Infection Tag} (`Yes'/`No') to indicate its current security status. This classification enables the dynamic grouping of assets into security zones that are agnostic to their physical network location, facilitating policy enforcement that follows the workload rather than the network interface.

Based on these classifications, the network is partitioned into distinct logical segments designed to enforce the principle of least privilege. The \textit{Controller Segment} houses the master node, restricting its communication strictly to essential command interactions; this isolation ensures that a compromise in downstream robotic nodes does not propagate to the central command system. Simultaneously, the \textit{Safe Robots Segment} encompasses all robotic nodes verified as uncompromised. Devices within this segment retain standard communication privileges, allowing for necessary peer-to-peer data exchange and control signal reception required for normal swarm operations.

To mitigate the risk of lateral movement during a breach, a dedicated \textit{Malicious Robots Segment} functions as a digital quarantine zone. Devices flagged with a positive infection status are dynamically reallocated to this segment, where a strict ``Block All'' policy is enforced. This policy effectively locks the compromised asset into a zero-trust state, denying all inbound and outbound traffic—including ICMP, TCP, and UDP protocols—until the device is remediated and reclassified. To streamline the application of these granular rules across the fleet, reusable configuration templates (specifically the Controller, Robots, and Quarantined Robots templates) were defined to ensure consistent and rapid policy propagation across the infrastructure.

The infection tag is updated locally on the edge device, based on an intrusion detection system.  The development of intrusion detectors for Internet-of-Things devices with embedded processors is itself a challenging problem, due to computational limits, intrinsic device/software heterogeneity, operational complexity, and other factors.  In a sequence of prior works, we have developed both passive and active anomaly detectors for IoT- and other cyber-physical- systems, including robotic platforms (e.g. \cite{ourwork1}).  A central innovation in these efforts is the use of physical-world data (e.g. robot position) in addition to network and embedded device data, to  detect anomalies in a platform-independent way.  These intrusion detectors, or other commercial or public-domain capabilities, can be used in a modular way to set the infection tag.  Because our focus here is on applying microsegmentation dynamically, we do not present intrusion detection capabilities in detail.  We also focus on testing the microsegmentation function in isolation from the intrusion detector.

\subsection{Automated Threat Response Framework}
A critical component of this methodology is the development of a bespoke automation framework designed to interface directly with the ColorTokens Xshield utility. To eliminate the latency associated with manual administrator intervention during a security breach, a Python-based script was deployed on the Master Controller. This script leverages the platform's API to continuously monitor the security posture of managed assets and dynamically enforce segmentation policies. By automating the interaction with the orchestration layer, the framework ensures that compromised nodes are isolated from the operational network in near real-time, significantly reducing the window of opportunity for lateral movement.

The automation algorithm operates on a continuous five-minute polling cycle, executing a closed-loop workflow of discovery, analysis, and enforcement. Initially, the script executes an asset discovery routine, querying the API to fetch a live inventory of all managed devices and storing their metadata in a local database. It subsequently retrieves the current tag assignments for each asset, specifically analyzing the \textit{Application} and \textit{Infection} tags to determine their logical segment location. This continuous monitoring establishes a real-time state awareness that serves as the foundation for the system's responsive capabilities.  We note that the five-minute polling cycle is defined by the default update time of ColorToken's segmentation function; however, this parameter can be reset to a user-defined value.

An intrusion detector module can be merged with the automation algorithm in several ways.  For a stand-alone intrusion-detection application, an API can be leveraged to extract information during each automation cycle, which can be used to determine the infection tag.  Alternately, for open-source or homemade detectors, multi-threaded or container-based  software architectures can be used.   

Upon the detection of a threat indicator, the framework triggers an immediate segmentation update. The script modifies the infection flag of the compromised asset to \texttt{Yes} and invokes the \texttt{assignAssetTags} function to push this state change to the cloud platform. This action automatically migrates the device to the \textit{Malicious Robots} segment. Crucially, the system utilizes the ``Auto Push to Firewall'' feature, ensuring that the new ``block-all'' policy is propagated to the network Gatekeeper without human approval. Once the threat is mitigated, the script executes a recovery routine by resetting the infection flag, thereby returning the device to the \textit{Safe Robots} segment and restoring full operational connectivity.

\subsection{Experimental Procedure}

\begin{figure}[h]
    \centering
    \includegraphics[width=0.85\columnwidth]{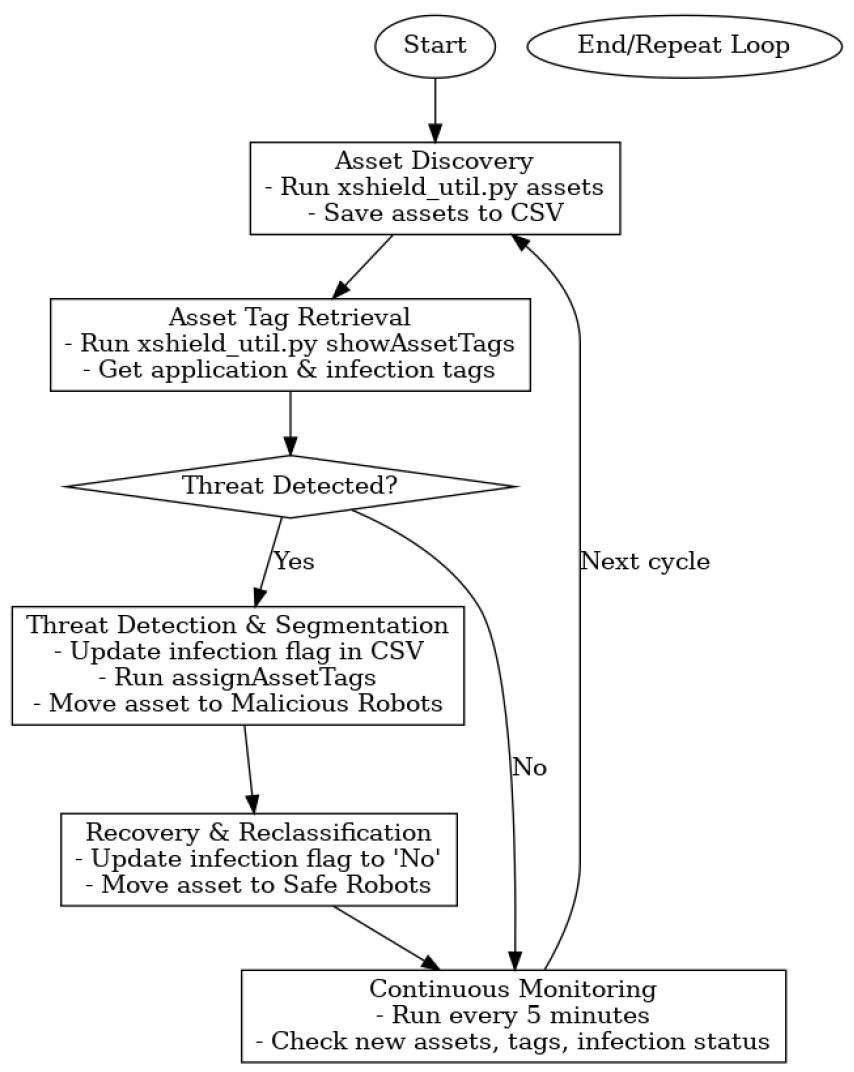}
    \caption{Zero-Trust Microsegmentation Automation Flowchart for Networked Robot Application.}
    \label{fig:2}
\end{figure}

Fig.~\ref{fig:2} explains that the automation algorithm operates in a continuous loop and follows these steps: 
\begin{enumerate}
    \item \textbf{\textit{Asset Discovery}}: (i) The script executes xshield\_util.py with the assets argument. (ii) All managed devices/assets are fetched and stored in a CSV file. (iii) This file serves as a live inventory for further monitoring and tagging.
    \item \textbf{\textit{Asset Tag Retrieval}}: (i) The script then calls the utility with the \texttt{showAssetTags} argument. (ii) This retrieves the current tags assigned to each asset, including application tags and infection status flags. (iii) These tags help determine whether an asset belongs to a ``safe'' or ``malicious'' segment.
    \item \textbf{\textit{Threat Detection and Segmentation}}: (i) When a cyber-threat is detected (via monitoring or external input), the corresponding asset is selected from the CSV file. (ii) The infection flag is updated in the CSV to reflect the new state (``Yes'' for infected). (iii) The script invokes xshield\_util.py with the \texttt{assignAssetTags} argument to update the asset’s segmentation. (iv) The infected device is automatically moved into a \textit{Malicious Robots} segment, effectively isolating it from safe devices.
    \item \textbf{\textit{Recovery and Reclassification}}: (i) Once the threat has been mitigated, the infection flag in the CSV is updated back to ``No''. (ii) The device is reclassified and moved back to the \textit{Safe Robots} segment. (iii) This dynamic reassignment ensures assets return to normal operation once verified as clean.
    \item \textbf{\textit{Continuous Monitoring}}: (i) The script runs on a 5 minute polling cycle. (ii) Every interval, it checks for: new or updated assets, tag changes, and infection status updates. (iii) This guarantees that segmentation policies remain aligned with the latest threat intelligence and allows for the policy changes to be applied.
\end{enumerate}

\begin{figure}[h]
    \centering
    \includegraphics[width=0.6\columnwidth]{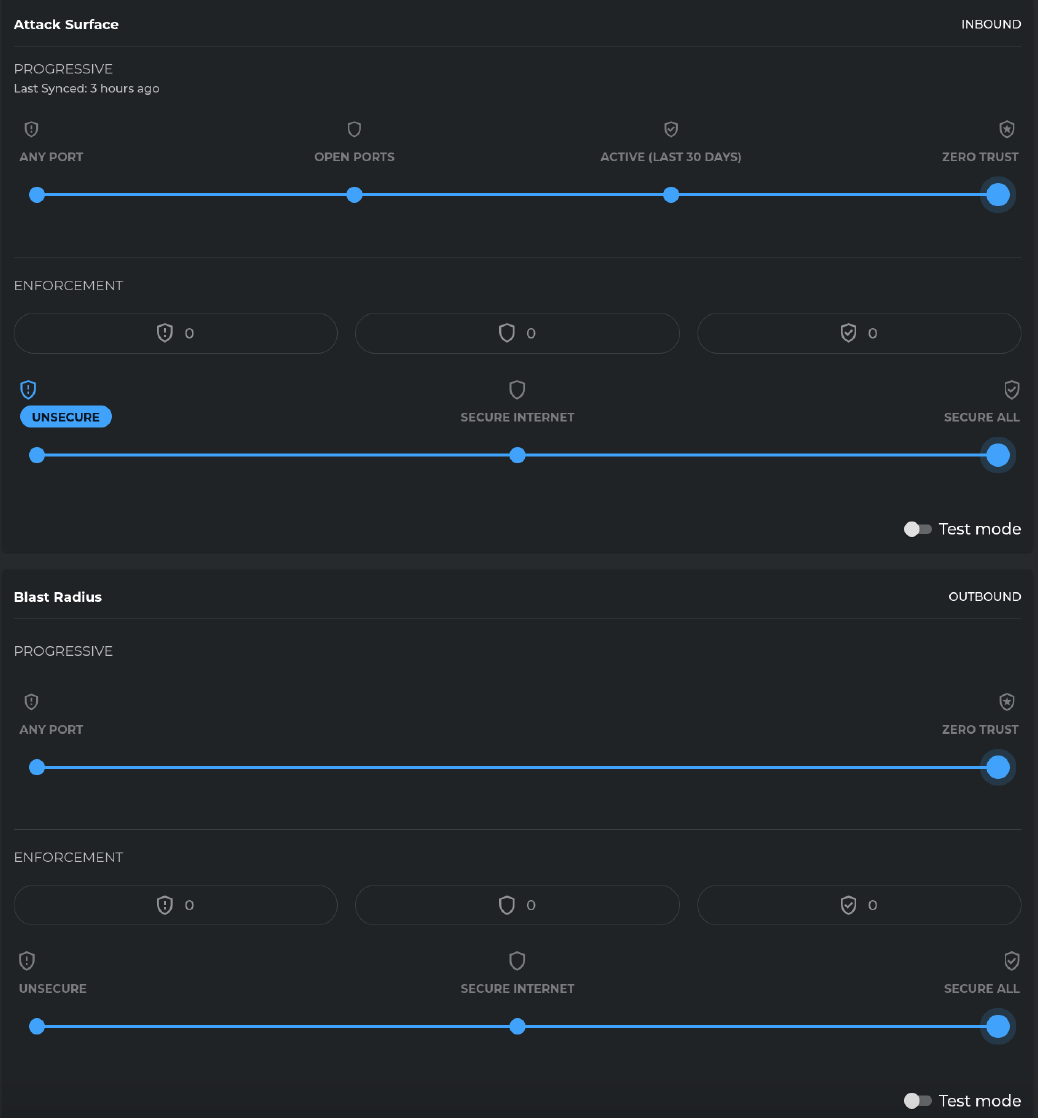}
    \caption{Policy Automation for Microsegmentation Implementation.}
    \label{fig:3}
\end{figure}

One of the features leveraged in this workflow is the Xshield ``Auto Push to Firewall'' policy automation option under the Segmentation settings (Fig.~\ref{fig:3}). When enabled, any tag or segment updates are automatically enforced on the firewall, ensuring immediate policy propagation. This provides near real-time zero-trust enforcement for infected devices without requiring manual administrator intervention.

\section{Results}
The experimental validation of the microsegmentation framework was conducted in three distinct phases: baseline traffic profiling, automated attack simulation with isolation, and post-breach recovery. The following subsections detail the quantitative observations and system behaviors recorded during these phases.

\subsection{Baseline Network Performance}
\begin{figure}[h]
    \centering
    \begin{subfigure}[b]{0.49\columnwidth}
        \centering
        \includegraphics[width=\linewidth]{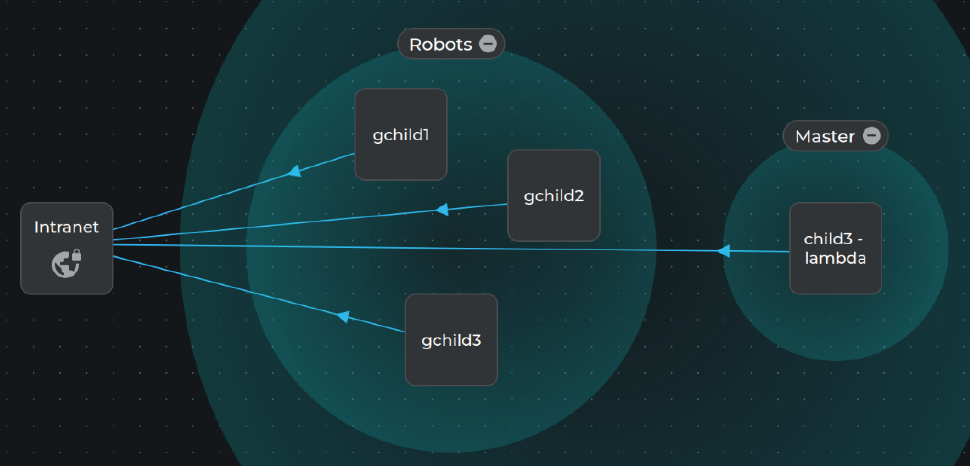}
        \caption{By Default.}
        \label{fig:4-1}
    \end{subfigure}%
    \hfill
    \begin{subfigure}[b]{0.49\columnwidth}
        \centering
        \includegraphics[width=\linewidth]{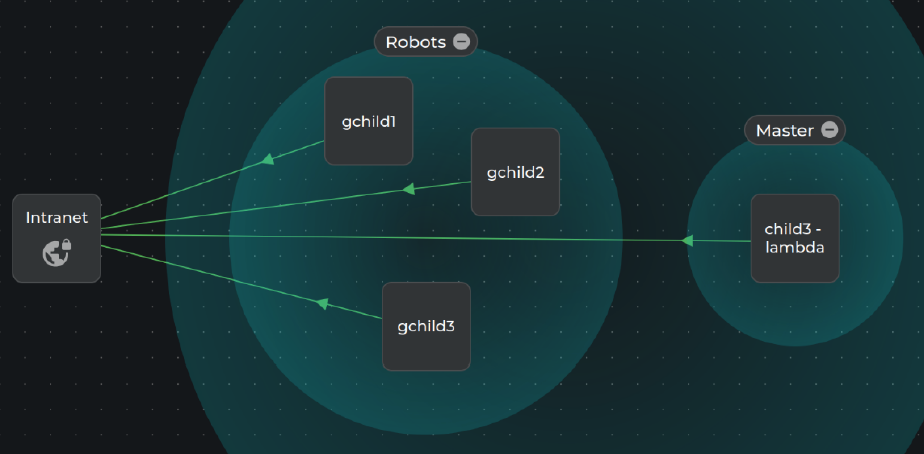}
        \caption{By Policy.}
        \label{fig:4-2}
    \end{subfigure}
    \caption{Initial Assets Allowed Connectivity.}
    \label{fig:initial-connectivity}
\end{figure}

Prior to the introduction of threat vectors, a baseline traffic profile was established to verify the operational integrity of the testbed. All devices, including the Master Controller and three robotic nodes, were onboarded and segmented according to their application roles. Analysis of the network traffic profile indicated steady, bidirectional communication between the controller and the robot nodes with no anomalous patterns. The deployment of the ColorTokens Gatekeeper and the enforcement of initial ``Safe Robot'' policies resulted in no observable performance degradation, ensuring that the security overhead did not impact the real-time requirements of the robotic control loops. Fig.~\ref{fig:4-1} shows the initial assets allowed connectivity by default, and Fig.~\ref{fig:4-2} shows the initial assets allowed connectivity by policy.

\subsection{Automated Isolation and Blast Radius Reduction}
To evaluate the efficacy of the automated response mechanism, specific robotic nodes (specifically \texttt{gchild1}, \texttt{gchild2}, and \texttt{gchild3}) were flagged as ``infected'' within the system. The automation script successfully detected the tag update and triggered a segment reassignment, moving the compromised devices from the \textit{Safe Robots} segment to the \textit{Malicious Robots} segment.

\begin{figure}[h]
    \centering
    \begin{subfigure}[b]{0.5\columnwidth}
        \centering
        \includegraphics[width=\linewidth]{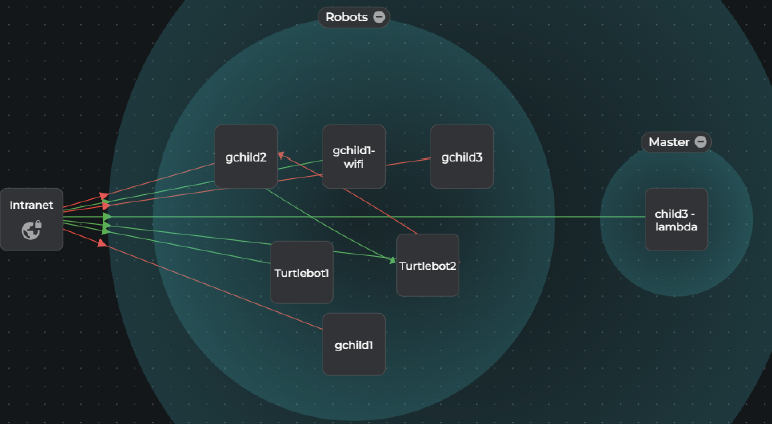}
        \caption{Attack Surface.}
        \label{fig:5-1}
    \end{subfigure}%
    \begin{subfigure}[b]{0.5\columnwidth}
        \centering
        \includegraphics[width=\linewidth]{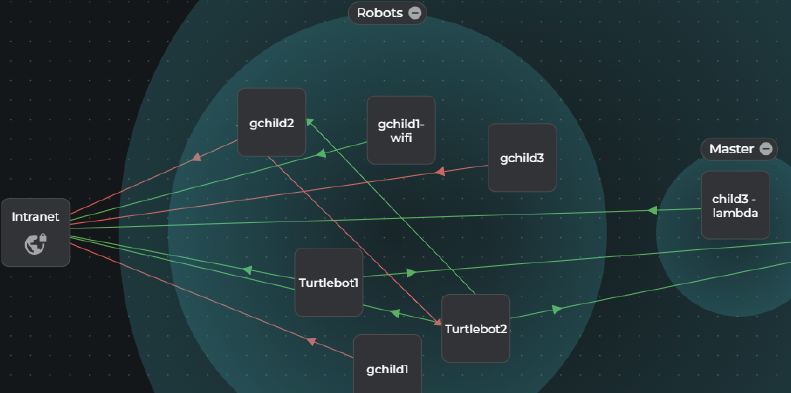}
        \caption{Blast Radius.}
        \label{fig:5-2}
    \end{subfigure}
    \caption{Attack Surface and Blast Radius after \texttt{gchild1}, \texttt{gchild2}, and \texttt{gchild3} are determined to be infected.}
    \label{fig:after-infection}
\end{figure}

System logs confirmed that the Xshield API pushed the updated firewall rules to the edge gateway within a single policy cycle. Visual analysis of the network topology (the ``Blast Radius'') demonstrated that while the infected nodes were quarantined, the Master Controller and remaining healthy assets continued to operate within their trusted zones, effectively preventing lateral movement. 
% Fig.~\ref{fig:5-1} and Fig.~\ref{fig:5-2} shows the attack surface and blast radius after gchild1, gchild2 and gchild3 is determined to be infected.

\subsection{Verification of Traffic Blockage}
\begin{figure}[h]
    \centering
    \includegraphics[width=0.9\columnwidth]{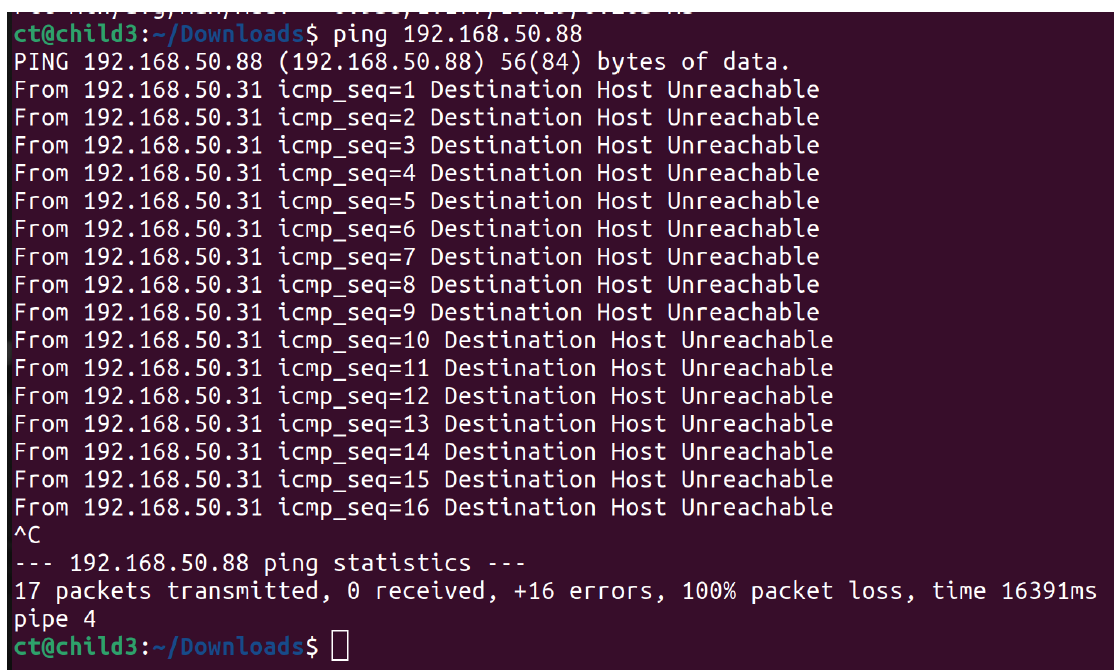}
    \caption{Controller trying to connect with the Infected Robot, the ICMP and SSH traffic is denied.}
    \label{fig:6}
\end{figure}

Quantitative verification of isolation was performed using active connectivity tests. Immediately following the quarantine action, the Master Controller attempted to initiate communication with an infected node via ICMP (ping) and SSH. As detailed in the system logs and illustrated in Fig.~\ref{fig:6}, the controller received ``Destination Host Unreachable'' responses for all transmission attempts. Statistical analysis of the traffic showed 100\% packet loss (17 packets transmitted, 0 received) for the infected target. This confirmed that the Zero-Trust policy successfully blocked all inbound and outbound traffic (ICMP, TCP, and UDP) for the compromised device, eliminating the possibility of Command and Control (C2) data exfiltration.

\subsection{Post-Incident Recovery}
The final experimental phase assessed the system's ability to restore operations following threat mitigation. Upon clearing the infection flag in the asset database, the automation detected the state change and reclassified the device back into the \textit{Safe Robots} segment. Observations confirmed that normal communication privileges were restored immediately without requiring a system reboot or manual firewall reconfiguration by an administrator. This verified that the dynamic segmentation model supports both rapid containment and efficient operational recovery. 
% Fig.~\ref{fig:7-1} and Fig.~\ref{fig:7-2} shows the network attack surface and blast radius in the process of recovering by changing the infected tag and updating the segments.

\begin{figure}[h]
    \centering
    \begin{subfigure}[b]{0.5\columnwidth}
        \centering
        \includegraphics[width=\linewidth]{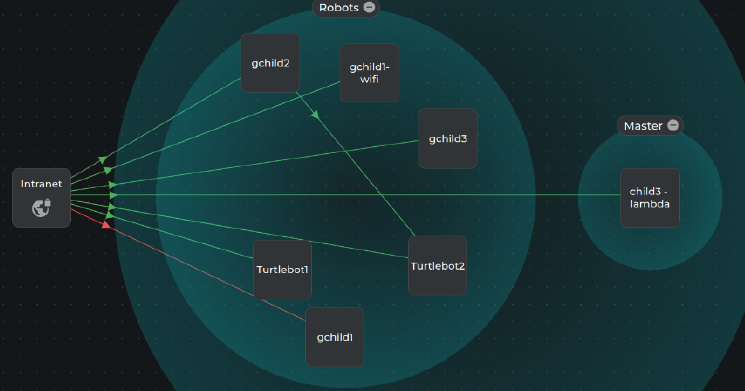}
        \caption{Attack Surface.}
        \label{fig:7-1}
    \end{subfigure}%
    \begin{subfigure}[b]{0.5\columnwidth}
        \centering
        \includegraphics[width=\linewidth]{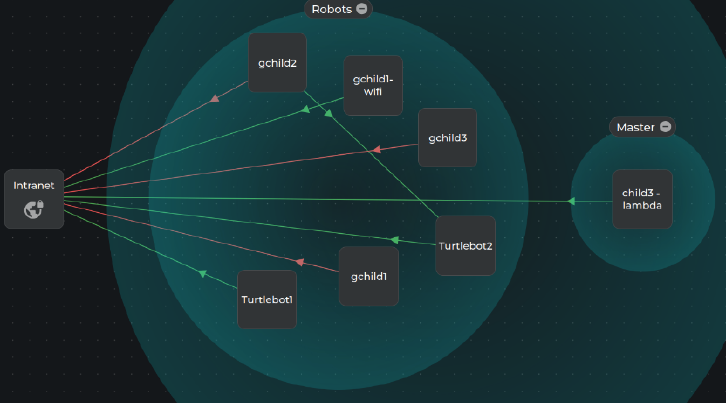}
        \caption{Blast Radius.}
        \label{fig:7-2}
    \end{subfigure}
    \caption{The network Attack Surface and Blast Radius during the recovery process by changing the infected tag and updating the segments.}
    \label{fig:recovery}
\end{figure}

% \newpage
\section{Future Work}
Future work will extend the demonstrated microsegmentation framework to highly dynamic cyber-physical systems, including chemical processes, drone swarms and autonomous vehicle fleets, where mobility and fluctuating node membership require adaptive policy scaling. Research efforts will also focus on integrating edge computing to minimize enforcement latency and incorporating AI-driven anomaly detection to enable fully-adaptive real-time threat isolation. Furthermore, we will consider adapting these zero-trust principles to cloud-native architectures with IoT endpoints, such as Docker/Kubernetes-based command and control systems.

\section{Conclusion}
This work demonstrated the practical feasibility of dynamic, behavior-adaptive microsegmentation for securing multi-robot and IoT networks. Using a testbed of three Raspberry Pi-powered Turtlebot nodes orchestrated via a commercial Zero-Trust platform, we showed that integrating real-time asset behavior monitoring with automated policy enforcement enables rapid containment of compromised devices while preserving the availability of healthy nodes. The closed-loop pipeline achieved sub-cycle isolation with 100\% blockage of ICMP, TCP, and UDP traffic to quarantined nodes and restored connectivity upon reclassification without observable overhead on the unaffected fleet. Beyond this demonstration, the study contributes an architectural pattern that decouples tag-based classification and API-driven orchestration from the underlying intrusion detector.

Several limitations scope these claims: the evaluation involves only three nodes in a controlled environment, and the intrusion detector was excluded to characterize the segmentation layer in isolation. The testbed provides a foundation for systematically measuring resilience metrics. Future work will extend these mechanisms to larger and more heterogeneous fleets, incorporate cyber-physical indicators for trust assessment, and integrate microsegmentation with federated anomaly detection for adaptive, low-latency threat isolation in mobile cyber-physical deployments.

\section*{Acknowledgment}
The authors extend their gratitude to ColorTokens for providing the Xshield enterprise platform and technical resources instrumental to this microsegmentation study, and also providing financial support for the work.

\begingroup
\footnotesize
\setlength{\itemsep}{0pt}
\setlength{\parskip}{0pt}
\renewcommand{\baselinestretch}{0.95}

\endgroup


\begin{thebibliography}{00}

\bibitem{iot1} Abuserrieh, Lobna, and Manar H. Alalfi. ``A survey on verification of security and safety in IoT systems.'' IEEE Access 12 (2024): 138627-138645.

\bibitem{iot2} Alaba, Fadele Ayotunde, et al. ``Internet of Things security: A survey.'' Journal of network and computer applications 88 (2017): 10-28.

\bibitem{iot3} Neshenko, Nataliia, et al. ``Demystifying IoT security: An exhaustive survey on IoT vulnerabilities and a first empirical look on Internet-scale IoT exploitations.'' IEEE Communications Surveys \& Tutorials 21.3 (2019): 2702-2733.

\bibitem{iot4} Yaacoub, Jean-Paul A., et al. ``Cyber-physical systems security: Limitations, issues and future trends.'' Microprocessors and microsystems 77 (2020): 103201.

\bibitem{iot5} Buck, Christoph, et al. ``Never trust, always verify: A multivocal literature review on current knowledge and research gaps of zero-trust.'' Computers \& Security 110 (2021): 102436.

\bibitem{b1} Humayed, Abdulmalik, Jingqiang Lin, Fengjun Li, and Bo Luo. ``Cyber-physical systems security—A survey.'' IEEE Internet of Things Journal 4, no. 6 (2017): 1802-1831.

\bibitem{zta1} Stafford, V. ``Zero trust architecture.'' NIST special publication 800, no. 207 (2020): 800-207.

\bibitem{zta2} He, Yuanhang, Daochao Huang, Lei Chen, Yi Ni, and Xiangjie Ma. ``A survey on zero trust architecture: Challenges and future trends.'' {\em Wireless Communications and Mobile Computing}, no. 1 (2022): 6476274.

\bibitem{cps1} Hahn, Adam, Roshan K. Thomas, Ivan Lozano, and Alvaro Cardenas. ``A multi-layered and kill-chain based security analysis framework for cyber-physical systems.'' {\em International Journal of Critical Infrastructure Protection} 11 (2015): 39-50.

\bibitem{cps2} Hassan, Wan Haslina. ``Current research on Internet of Things (IoT) security: A survey.'' Computer networks 148 (2019): 283-294.

\bibitem{b2} Baligodugula, Vishnu Vardhan, Ashutosh Ghimire, and Fathi Amsaad. ``An overview of secure network segmentation in connected IIoT environments.'' Computing\&AI Connect 1, no. 1 (2024): 1-10.

\bibitem{inter1} Arifeen, Murshedul, Andrei Petrovski, and Sergei Petrovski. ``Automated microsegmentation for lateral movement prevention in industrial internet of things (IIoT).'' In 2021 14th international conference on security of information and networks (SIN), vol. 1, pp. 1-6. IEEE, 2021.

\bibitem{inter2} Jensen, Michael, Aisha Rahman, Carlos Vega, Emily Zhao, and Ravi Chandrasekar. ``Zero Trust Architectures for Securing Cyber-Physical Systems: A Comprehensive Security Framework for Critical Infrastructure Protection.''

\bibitem{inter3} Osman, Amr, Armin Wasicek, Stefan K\"{o}psell, and Thorsten Strufe. ``Transparent microsegmentation in smart home {IoT} networks.'' In 3rd USENIX Workshop on Hot Topics in Edge Computing (HotEdge 20). 2020.

\bibitem{intrud1} Chaabouni, Nadia, Mohamed Mosbah, Akka Zemmari, Cyrille Sauvignac, and Parvez Faruki. ``Network intrusion detection for IoT security based on learning techniques.'' IEEE Communications Surveys \& Tutorials 21, no. 3 (2019): 2671-2701.

\bibitem{intrud2} Smys, Smys, Abul Basar, and Haoxiang Wang. ``Hybrid intrusion detection system for internet of things (IoT).'' Journal of ISMAC 2, no. 04 (2020): 190-199.

\bibitem{intrud3} Li, Jiaqi, Zhifeng Zhao, Rongpeng Li, and Honggang Zhang. ``Ai-based two-stage intrusion detection for software defined iot networks.'' IEEE internet of Things Journal 6, no. 2 (2018): 2093-2102.

\bibitem{b3} Sarhan, Mohanad, Wai Weng Lo, Siamak Layeghy, and Marius Portmann. ``HBFL: A hierarchical blockchain-based federated learning framework for collaborative IoT intrusion detection.'' Computers and Electrical Engineering 103 (2022): 108379.

\bibitem{b4} S\'{a}ez-de-C\'{a}mara, Xabier, Jose Luis Flores, Crist\'{o}bal Arellano, Aitor Urbieta, and Urko Zurutuza. ``Clustered federated learning architecture for network anomaly detection in large scale heterogeneous IoT networks.'' Computers \& Security 131 (2023): 103299.

\bibitem{b5} Nguyen, Van Tuan, and Razvan Beuran. ``FedMSE: Federated learning for IoT network intrusion detection.'' arXiv e-prints (2024): arXiv-2410.

\bibitem{b6} Abd Elaziz, Mohamed, Ibrahim A. Fares, Abdelghani Dahou, and Mansour Shrahili. ``Federated learning framework for IoT intrusion detection using tab transformer and nature-inspired hyperparameter optimization.'' Frontiers in Big Data 8 (2025): 1526480.

\bibitem{ourwork1} Ajeigbe, Oluwafemi, Jaewon Kim, Ryan Ozelton, Jeremy Tang, Anthony Munoz, and Sandip Roy. ``Strengthening Cyber Defenses for Networked Autonomous Robots.'' In 2025 IEEE 21st International Conference on Automation Science and Engineering (CASE), pp. 2402-2407. IEEE, 2025.

\bibitem{ourwork2} Kim, Jaewon, and P. R. Kumar. ``Security of control systems with erroneous observations.'' IFAC-PapersOnLine 53, no. 2 (2020): 2225-2230.

\bibitem{b7} Lam, An Ngoc, Oystein Haugen, and Jerker Delsing. ``Dynamical orchestration and configuration services in industrial iot systems: An autonomic approach.'' IEEE Open Journal of the Industrial Electronics Society 3 (2022): 128-145.

\bibitem{b8} Tahir, Nazish, and Ramviyas Parasuraman. ``Edge computing and its application in robotics: A survey.'' Journal of Sensor and Actuator Networks 14, no. 4 (2025): 65.

\end{thebibliography}
\end{document}